\documentclass[aps,english,10pt,superscriptaddress,twocolumn]{revtex4}
\usepackage{amsfonts}
\usepackage{bbm}
\usepackage{epsfig}
\usepackage{times}
\usepackage{babel}
\usepackage{color}
\usepackage{hyperref}
\usepackage{framed}
\usepackage{changes}
\usepackage{physics}
\usepackage{mathtools}
\usepackage{amsthm}
\usepackage{amsmath}
\usepackage{amssymb}
\usepackage{tensor}
\usepackage{graphicx}
\usepackage{float}
\usepackage{color}

\DeclareGraphicsExtensions{.png, .pdf}
\graphicspath{{./img/}}

\newcommand{\qexp}[1]{\langle #1 \rangle}

\begin{document}

\title{Estimates of the Quantum Fisher Information in the $S=1$ Anti-Ferromagnetic Heisenberg Spin Chain with Uniaxial Anisotropy}
\date{\today}
\author{J. Lambert}
\email{lambej3@mcmaster.ca}
\author{E. S. S{\o}rensen}
\email{sorensen@mcmaster.ca}
\affiliation{Department of Physics \& Astronomy, McMaster University
1280 Main St.  W., Hamilton ON L8S 4M1, Canada}

\begin{abstract}
The quantum Fisher information is of considerable interest not only for quantum metrology
but also because it is a useful entanglement measure for finite temperature mixed states. In particular,
it estimates the degree to which multipartite entanglement is present. Recent results have
related the quantum Fisher information to experimentally measurable probes. While in principle possible,
a direct evaluation of the quantum Fisher information at finite temperatures is technically challenging
and here we show that a simple estimate can be obtained for materials where the single mode approximation
is valid.  We focus on the $S=1$ anti-ferromagnetic Heisenberg model with uniaxial anisotropy.  
Quantum Monte Carlo thechniques are  used to determine low temperature correlations from which the quantum Fisher
information can be estimated within the single mode approximation.  The quantum Fisher information is compared to the quantum variance for the
staggered magnetization operators in the transverse direction and inequalities between the quantum Fisher information, the quantum variance and the full variance are discussed.
Both the quantum and full variance as well as the quantum Fisher information are examined at finite
temperatures above the isotropic point and at the quantum critical 
point for the Haldane-N\'{e}el transtion. A finite size scaling study of the quantum
Fisher information is performed at the quantum critical point and used to confirm the Ising nature of the
Haldane-N\'{e}el transition.
\end{abstract} 
\maketitle

\section{Introduction}\label{sec:introduction}

The quantum Fisher information (QFI), $F_Q$, is often studied in quantum metrology~\cite{Petz1996,Petz2002,Paris2009,Toth2012,Toth2013,Toth2014}.
There, one considers unitary dynamics $U=\exp(-i \hat{O}\theta)$ and the phase estimation sensitivity is then limited by the Cram\'er-Rao bound $(\Delta\theta)^2\geq 1/F_Q[\hat{O}]$
for {\it any} measurement. From a condensed matter perspective, the quantum Fisher information is particularly interesting since it
can be used to estimate multipartite entanglement even at finite temperatures
since $F_Q/N>m$ with $m$ a divisor of $N$ signals ($m$+1)-partite entanglement~\cite{Pezze2009,hyllus2012fisher,Toth2012,Toth2014}. 
Significant progress in the understanding of, in particular
bi-partite, entanglement in quantum many-body systems
has been made\cite{eisert2010colloquium,Laflorencie2016,headrick2010entanglement}.
More recently, a host of
techniques have been developed to efficiently quantify  multipartite
entanglement in quantum many-body systems. (For a review of
entanglement witnesses see \cite{horodecki2009quantum, vedral1997quantifying,
amico2008entanglement, plenio2005introduction}). 
For our purpose we will take the definition of multipartite entanglement to be
the natural generalization of bipartite entanglement. Namely, consider and
$N$-body quantum state $\ket{\psi_N}$. Now imagine expressing this state as a
product of $m$ states each containing $N_m$ particles $\ket{\psi_N} =
\bigotimes_{i=0}^m\ket{\phi_i}$. A $k$-partite entangled state is one for which
the largest constituent state $\phi_i$ contains $N_i=k$ particles, and cannot
be further decomposed. It's clear that one can recover from this the usual
definition of bipartite entanglement. 
Ideally, for the study of multipartite entanglement, one would like to use
techniques that do not rely on a particular knowledge of the density matrix, as
these are the techniques most easily connected to experiment and the quantum Fisher information
seem well suited for this purpose.

The quantum Fisher information (QFI) has long been known as a monotonic
multipartite entanglement measure \cite{Pezze2009,hyllus2012fisher,Toth2012,Toth2014}, but only recently
has it been connected to the dynamic structure factor which is easily accessed
by experimental probes such as neutron scattering \cite{NaturePhysics}. This has led
to the studies of the QFI and multipartite entanglement in the Kitaev chain~\cite{Pezze2017}, quantum Ising chain~\cite{NaturePhysics},
XY spin chain~\cite{Liu2013}, XXZ spin chain~\cite{Zheng2015} and Lipkin-Meshkov-Glick model~\cite{Li2013,Ma2009}. In order to access the QFI
these studies all rely on the exact solvability of the models considered and
from a numerical perspective, accessing the Fisher information can be challenging 
in particular at finite temperature for realistic non-integrable quantum many-body models.
Here we show that a  simple estimate of the QFI, $F_Q^\text{SMA}$, can be obtained by using
the single mode approximation (SMA) which allows the QFI to be calculated directly from the
equal time structure factor. The quantum variance (QV), has been established as a lower bound
for $F_Q$,  $F_Q \geq 4\qexp{\delta^2\hat{O}}_Q$ \cite{frerot2016quantum} and at the same
time an upper bound is given by the full variance $F_Q \leq 4\qexp{\delta^2\hat{O}}$~\cite{Toth2014}. 
This then serves as a  rigorous check on the validity of the SMA calculations.

We focus on the  $S=1$ AFM Heisenberg model with uniaxial anisotropy,
\begin{equation}
	\hat{H} = J \sum_i \left(\boldsymbol{S}_i\cdot \boldsymbol{S}_{i+1} + D(S_i^z)^2\right),
\end{equation}
where $D$ is the uniaxial anisotropy and we shall take $J=1$ throughout. At $D=0$ this model displays the celebrated Haldane gap
at $k=\pi$ of $\Delta\sim0.41J$ and it is quite well established~\cite{so1994equal,Golinelli1999} that
the single mode approximation works very well around $k=\pi$ for moderate values of $D$.
We perform stochastic series expansion~\cite{sandvik1991quantum,sandvik1998stochastic,Syljuaasen2002} (SSE) quantum Monte Carlo simulations
to evaluate low-temperature equal time correlations, from which $F_Q^\text{SMA}$ is obtained, as well as finite temperature calculations
to determine the quantum and full variance. This demonstrates the presence of significant multi-partite entanglement even at the isotropic point
$D=0$.

The layout of this paper is as follows. In Section~\ref{sec:techniques} we introduce some of the
key properties of the QFI and QV (section~\ref{sec:qfi}). We then introduce the single mode
approximation (section~\ref{sec:sma}) and its application to the  $S=1$ AFM Heisenberg model. Then in section~\ref{sec:results}
we present SSE results for the system QV and the QFI at the isotropic point 
as well as for a range of values $D<0$ towards the quantum critical
point before turning to our conclusions in section~\ref{sec:conclusions}.

\section{Techniques}\label{sec:techniques}

\subsection{QFI and QV}\label{sec:qfi}

The quantum Fisher information is one possible generalization of the classical
Fisher information, which quantifies the distinguishability of a family of
distributions parameterized by one (or possibly several) parameters
$\theta$\cite{wootters1981statistical,braunstein1994statistical}. The
quantum generalization of this quantifies the distinguishability of a family of
quantum states defined by,
\begin{equation}
  \rho(\theta)=e^{-i\theta\hat{O}}\rho e^{i\theta\hat{O}},
\end{equation}
where $\hat{O} = \sum_r \hat{O}^\alpha$ is a sum over local operators. In particular, the QFI
can be thought of as the statistical speed related to the rate of change of the
Bures' distance, which is a metric on the space of density matrices
\cite{gessner2017statistical}. 
For a density matrix that in its eigenbasis is given by:
\begin{equation}
  \rho=\sum_\lambda p_\lambda |\lambda\rangle\langle \lambda|,
\end{equation}
the QFI is given by, 
\begin{equation}
 F_Q = 2\sum_{\lambda,\lambda'}\frac{(p_\lambda - p_{\lambda'})^2}{p_\lambda+p_{\lambda^\prime}}\qty|\langle \lambda^\prime |\hat{O}|\lambda\rangle|^2.
	\label{eq:finitetqfi}
\end{equation}
The relationship between the QFI and the multipartite entanglement has been
well established in\cite{Pezze2009,hyllus2012fisher,Toth2012,Toth2013,Toth2014}. In particular, for a QFI density:
\begin{equation}
f_Q \equiv F_Q / N > m,
\end{equation} where $m$ is a divisor of $N$, the system is $(m+1)$-partite
entangled. The QFI thus increases monotonically with the entanglement. One of
the most appealing features of the QFI is that it is defined for mixed states,
allowing one to determine the entanglement content of a state at finite
temperature. Recent work~\cite{NaturePhysics} has connected the QFI density to the dynamic structure factor, 
\begin{equation}
	f_Q(k) = \frac{2}{N^d\pi}\int_{-\infty}^{\infty}\dd \omega \tanh^2\left(\frac{\omega}{2T}\right)S(\omega, k).
	\label{eq:ftqfidensity}
\end{equation}
The dynamic structure factor is routinely measured in inelastic neutron
scattering experiments and thus provides a highly accessible measure of the
multipartite entanglement of a system. 
In the zero temperature limit the QFI Eq.~(\ref{eq:finitetqfi}) reduces to the variance of the
operator $\hat{O}$,
\begin{equation}
	F_Q = 4\left(\langle\hat{O}^2\rangle - \langle\hat{O}\rangle^2\right)
	\label{eq:zerotqfi}
\end{equation}

Another experimentally accessible entanglement monotone is the quantum
variance\cite{frerot2016quantum}. The idea is that at finite temperature
both thermal and quantum fluctuations contribute to the variance, 
\begin{equation}
  \qexp{\delta^2\hat{O}}\equiv\qexp{\hat{O}^2} - \qexp{\hat{O}}^2,\label{eq:Vartot}
\end{equation}
so that we may write
\begin{equation}
\qexp{\delta^2\hat{O}}=\qexp{\delta^2\hat{O}}_Q + \qexp{\delta^2\hat{O}}_T,
\end{equation}
with the quantum fluctuations being some indicator of the extent to which a
state may be entangled. 
In order to isolate the quantum component of the
fluctuations we may use the fact that the thermal component of the fluctuations
is simply given by the susceptibility. We therefore have,
\begin{equation}
	\langle\delta^2\hat{O}\rangle_Q = \langle\delta^2\hat{O}\rangle - \chi_{\hat{O}}k_B T.
\end{equation}
It can be shown that the QV is actually a lower bound on the QFI via the relation \cite{frerot2016quantum},
\begin{equation}
	 4\langle\delta^2\hat{O}\rangle_Q\leq F_Q.
\end{equation}
Additionaly we can see that the total variance of the operator must be an upper bound to the QFI 
\cite{Pezze2017}. In section III. we compute these quantities and explicitly show that they serve as a hard upper and lower bound
to the single-mode approximated QFI. 
Both the QFI and the QV are thought to take a universal form at the quantum
critical point. The exact scaling behaviour of these quantities will ultimately
be inherented from the operator in terms of which they are defined. 

The work in \cite{NaturePhysics} derives the scaling exponents for the QFI
density at both zero and finite temperature. We summarize their results here
for convenience. For a review of scaling theory one can refer to
\cite{cardy1996scaling}. 
Consider a rescaling of the lattice by an amount $\lambda$. The operator
$\hat{O}$ will then rescale by some amount $\lambda^{-\Delta_\alpha}$. The QFI
density will therefore scale as $\lambda^{d - 2\Delta_\alpha}$. Thus, we can
identify $\Delta_Q = d - 2\Delta_\alpha$ as the scaling dimension for the QFI
density. This result holds in the finite temperature case as well. In order to
demonstrate this we recall that the temperature and frequency both scale with
the dynamical critical exponent $z$. By examining 
Eq.~(\ref{eq:ftqfidensity}), we see that the argument of the  hyperbolic tangent
function is thus scale invariant. That leaves us with the scaling of the
dynamical structure factor which scales in the same was as the correlation
function, and thus the finite temperature QFI will also scale as $\Delta_Q =
d-2\Delta_\alpha$. For large but finite systems at low but non-zero temperatures
we then expect~\cite{NaturePhysics}
\begin{equation}
  f_Q(T,L)=\lambda^{\Delta_Q} h(\lambda^z T,\lambda/L),
\end{equation}
where $L$ is the linear size of the system. If simulations are performed at low enough temperatures
that the scaling with $T$ can be neglected, it then follows from finite-size scaling that
\begin{equation}
  f_Q(L)\sim L^{\Delta_Q}.\label{eq:fscaling}
\end{equation}

\subsection{The Single Mode Approximation}\label{sec:sma}

We consider the first principles definition of the structure factor for the spectrum of the Hermitian operator $\hat{O}$ \cite{lovesey1980condensed}
\begin{equation}
	S(\omega, k) = 2\pi\sum_{\lambda,\lambda'}p_{\lambda}\qty|\bra{\lambda'}\hat{O}\ket{\lambda}|^2\delta(\omega + E_\lambda - E_{\lambda'}),
	\label{eq:sf}
\end{equation}
where $p_\lambda = e^{\beta E_\lambda}/\mathcal{Z}$. 
The structure factor is a function of $k$ through the definition of the $\hat{O}$. In the limit of $T\to 0$ it can be shown that Eq.~(\ref{eq:sf}) takes on the simpler form:
\begin{equation}
	S(\omega, k) = \sum_{i,\lambda'}\qty|\bra{\lambda'}\hat{O}\ket{0}_i|^2\delta(\omega + E_0 - E_{\lambda'}).
	\label{eq:zt}
\end{equation}
Here $\ket{0}$ is intended to represent the ground state. In general, the
ground state may be degenerate. The summation index $i$ includes all states
having the ground state energy $E_0$. The content of the single mode
approximation is twofold. First we assume that only the first two energy levels
are substantially populated. Second, we assume that transitions from the ground
states to state at energies above the first excited state have negligible
matrix elements compared with transition from the ground state manifold to the
first excited state. That is to say:
\begin{equation}
	S(\omega, k) = S_0(k)\delta\left(\omega - \omega_k^{(01)}\right) + \tilde{S}(k, \omega),
	\label{sma}
\end{equation}
where $\qty|S_0(k)| \gg \qty|\int_{-\infty}^{\infty}\tilde{S}(\omega,
k)\dd\omega|$, and $\omega_k^{(01)}  \coloneqq E_1 - E_0$. In other words, the
bulk of the spectral weight is on the transition between the ground state and
the first excited state. The $\tilde{S}(\omega, k)$ represents the spectral
weight coming from states above the first excited state. It is important to ask
how this approximation behaves at finite temperatures where the population of
excited states will increase. Let's consider the leading order correction to
Eq.~(\ref{eq:sf}) at finite temperatures. We write the partition function as
$\mathcal{Z} = \sum_i \eta_i \exp(-\beta E_i)$, where $\eta_i$ denotes the
degeneracy of the $i^{\text{th}}$ energy level. We may thus write the partition
function as, $\mathcal{Z}=\exp(-\beta E_0)\left(\eta_0 +
\sum_{\lambda}\eta_\lambda\exp(-\beta \omega^{(0\lambda)}_k)\right)$ where
$\omega_{0\lambda}$ is the energy difference between the ground sate and
$\ket{\lambda}$. Substituting this into Eq.~(\ref{eq:sf}) and taking the
temperature to be zero recovers Eq.~(\ref{eq:zt}). Taking instead the
temperature to be small but non-zero, the next leading order correction is
given by,

\begin{align}
	S(\omega, k) = &\frac{\delta\left(\omega-\omega^{(01)}_k\right) + \delta\left(\omega+\omega^{(01)}_k\right)e^{-\beta\omega^{(01)}_k}}				{\eta_0 + \eta_1 e^{-\beta\omega^{(01)}_k}}S_0(k) \nonumber \\
		&+ \tilde{S}(\omega, k).
\end{align}
Not surprisingly, the next leading order correction is spectral weight coming from transition from the first excited state to the ground state. In general when integrated over, this thermal correction will manifest as a multiplicative factor of the form,
\begin{equation}
	A = \frac{1 + e^{-\beta\omega_k^{(01)}}}{\eta_0+\eta_1e^{-\beta\omega_k^{(01)}}}.
	\label{eq:factor}
\end{equation}
When the ground state and excited state are non-degenerate, this factor will be equal to unity, and therefore will not affect the quantum Fisher information.
For a more thorough examination of the effects of temperature on the entanglement near a critical point see \cite{gabbrielli2018multipartite}. The approach there is effecitively a single-mode approximation but applied directly to the QFI and used to examine integrable systems.   
\begin{figure}
	\includegraphics[width=0.48\textwidth]{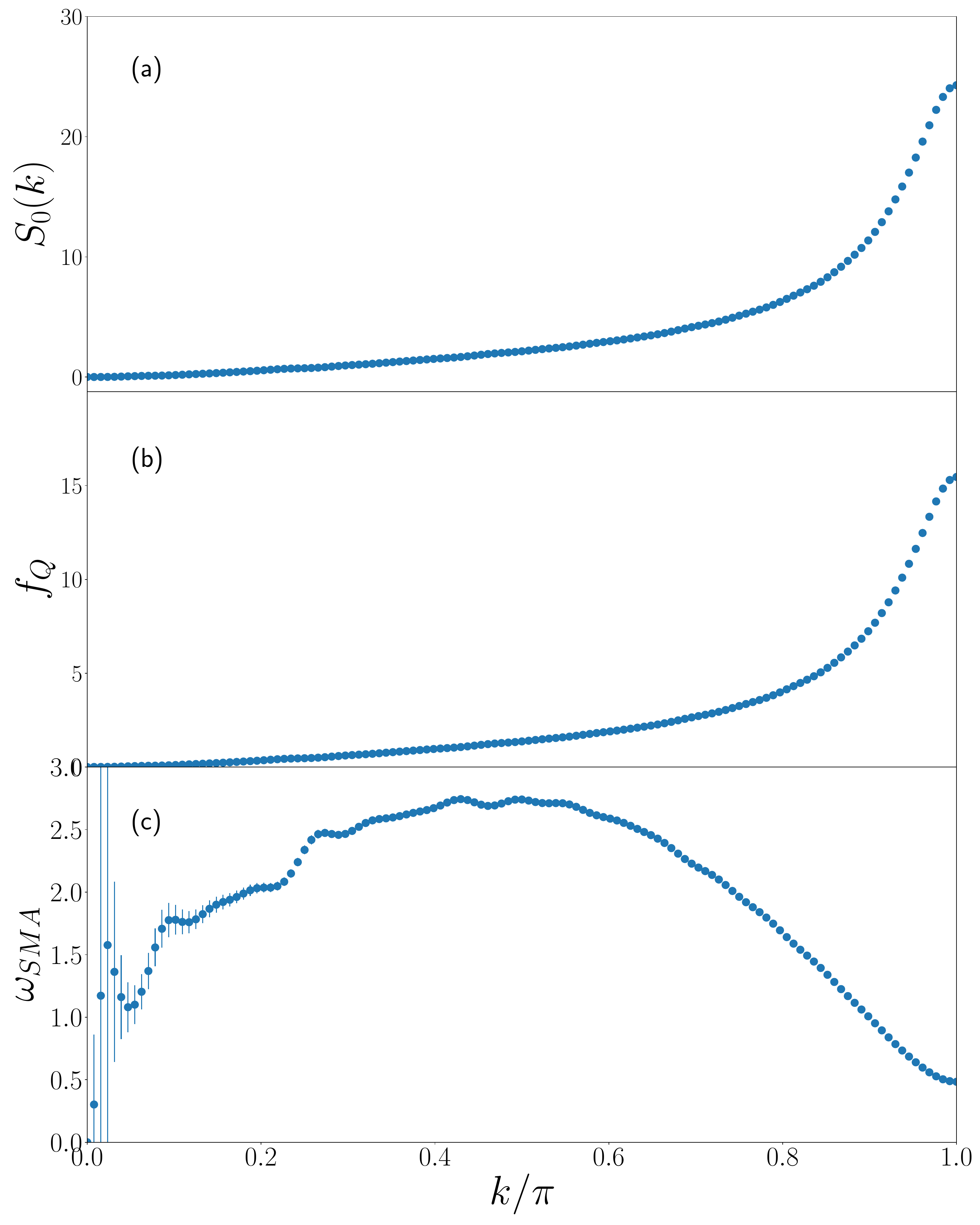}
        \caption{(Color online) (a) The equal time structure factor for the isotropic $S=1$ AFM
        model with $N=256$, $\beta=400$, exhibiting at peak at $k=\pi$.
        (b) The QFI density in the first Brillouin zone approaching
        zero at $k=0$, and exhibiting a peak at
        $k=\pi$. (c) $\omega_k$ as obtained from the single-mode approximation.
        The characteristic gap of $0.41J$ at $k=\pi$ is clearly visible.}
	\label{fig:isochar}
\end{figure}

In order to employ the single mode approximation we need some way to determine
the gap, $\omega^{(01)}_k$ (we henceforth drop the superscript and allow
$\omega_k$ to denote the dispersion for the first excited state).  It is clear
that due to energy conservation $\text{supp}(\tilde{S}) = \left\{\omega :
\omega_k < \omega_c < \omega\right\}$, where $\omega_c$ denotes the bottom of
the continuous portion of the energy spectrum. In order to determine $\omega_k$
we multiply Eq.~(\ref{sma}) by $\omega$ and integrate over all frequencies: 
\begin{equation}
	\frac{1}{\omega_k}\int_{-\infty}^{\infty}\dd\omega\omega S(\omega,k) = S_0(k) + 
	\frac{1}{\omega_k}\int_{-\infty}^{\infty}\dd\omega\omega \tilde{S}(\omega, k).
	\label{inter}
\end{equation}
In order to deal with the $\tilde{S}$, we note that:
\begin{equation}
	\frac{1}{\omega_k}\int_{-\infty}^{\infty}\dd\omega\omega\tilde{S}(\omega, k) \geq
	\int_{-\infty}^{\infty}\dd\omega \tilde{S}(\omega, k).
\end{equation}
This assertion is made valid by the positive semi-definite nature of $\tilde{S}$. By substituting this inequality into Eq.~(\ref{inter}) we see that the LHS is, by definition, the equal time structure factor $S(k)$, giving:
\begin{equation}
	\frac{1}{S(k)} \int_{-\infty}^{\infty}\dd\omega\omega S(\omega, k) \geq \omega_k,
\end{equation}
with
\begin{equation}
  S(k)\equiv\int S(k,\omega)d\omega=\sum_r e^{-ikr}\langle S(r)\cdot S(0)\rangle.
\end{equation}
We may use the following sum rule \cite{hohenberg1974sum}:
\begin{equation}
	\int_{-\infty}^{\infty} \dd \omega\omega S(\omega, k) = \pi\qexp{[\hat{O}^\dagger, [H, \hat{O}]]},
\end{equation}
to evaluate this expression, which leaves the bound on $\omega_k$ as:
\begin{equation}
  \omega_k \leq \omega_{\text{SMA}}(k) \coloneqq \pi\frac{\qexp{[\hat{O}^\dagger, [H, \hat{O}]]}}{S(k)}.
\label{disp}
\end{equation}
Here, $S(k)$ along with the different components of the commutator can relatively easily be estimated using
quantum Monte Carlo methods from which $\omega_{\text{SMA}}(k)$ can then be obtained.
Results are shown in Fig.~\ref{fig:isochar}(a) for the $S(k)$ from Fourier transforms of the ground-state correlation
functions obtained from QMC calculations. Fig.~\ref{fig:isochar}(c) shows the resulting $\omega_{SMA}(k)$.

\begin{figure}[t!]
	\includegraphics[clip, width=0.5\textwidth]{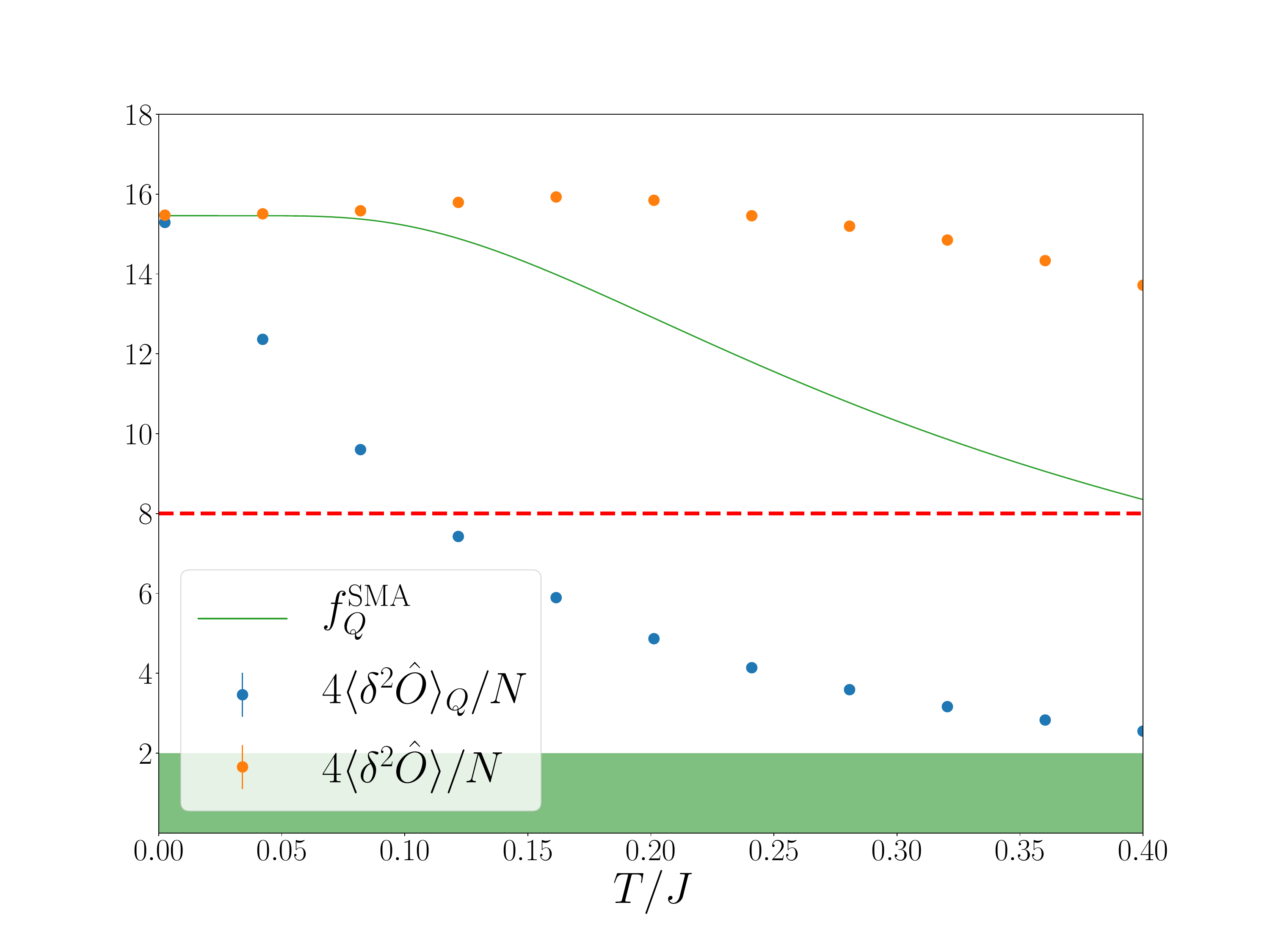}
        \caption{Finite temperature behaviour of the QFI density, $f_Q(k=\pi)$ and the
        quantum variance for temperatures up to the Haldane gap at the
        isotropic point, $D=0$, for $N=256$. 
        $f_Q^\text{SMA}$ is obtained from simulations at $\beta=400$.
Upper and lower bounds for $f_Q$ given by $4\langle\delta^2\hat{O}\rangle/N$ and $4\langle\delta^2\hat{O}\rangle_Q/N$ are also shown for a range of temperatures.
        The green shaded region indicates the level
        the $f_Q$ has to exceed to indicate the presence of {\it more than} bi-partite entanglement.
        The dashed red line indicates the threshold for (8+1)-partite entanglement. Below that line and above the green region the system would be
	(4+1)-partite entangled.}
	\label{fig:isotemp}
\end{figure}

\section{Results}\label{sec:results}
We now turn to a discussion of our results for the QFI and multipartite entanglement in 
the $S=1$ AFM Heisenberg model with uniaxial anisotropy ,
\begin{equation}
	\hat{H} = J \sum_i \left(\boldsymbol{S}_i\cdot \boldsymbol{S}_{i+1} + D(S_i^z)^2\right).
\end{equation}
This model has several appealing feature to investigate multipartite
entanglement. First, it possesses a symmetry-protected topologically (SPT)
phase with a gapped ground state in the isotropic region, with $\Delta \approx
0.41 J$\cite{haldane1983nonlinear, affleck1990theory}. This phase is
characterized by the breaking of a hidden $\mathbb{Z}_2\times\mathbb{Z}_2$
symmetry which establishes a long-range string order \cite{kennedy1992hidden}.
Second, the uniaxial anisotropy can drive two quantum phases transitions with
critical points falling into two different universality classes. The phase
diagram of this model has been extensively investigated in
\cite{zhang2013phase, albuquerque2009quantum, chen2003ground}. The first 
transition is from the topologically protected Haldane phase to a disordered phase
with quasi-N\'{e}el ordering ($D_C^{HN} \approx -0.31$). The second transition
is to a phase which is often called the "large-D phase"
($D_C^{HL}\approx 0.98$). This latter phase is essentially "empty" as the large
unixial anisotropy forces each spin to have zero $S^z$ projection. The
Haldane-N\'{e}el transition is in the universality class of the 2D Ising model,
while the Haldane-Empty transition is in the Gaussian universality class. The
excitation spectrum exactly at the isotropic point consists of a triplet state.
This degeneracy is lifted away from the isotropic point into a heavier magnon
with energy $\omega_k^{(\parallel)}$ and a lighter doublet with
$\omega_k^{(\bot)}$. This notation is meant to evoke the fact that the heavier
magnon is in the direction parallel to the uniaxial anisotropy, while the
doublet corresponds to the transverse excitations.  Most importantly, in a sizable region
around $k=\pi$, as well as for $D\neq0$, the dynamical strucure factor is well approximated by a single mode.

\begin{figure}[t!]
	\includegraphics[width=0.5\textwidth]{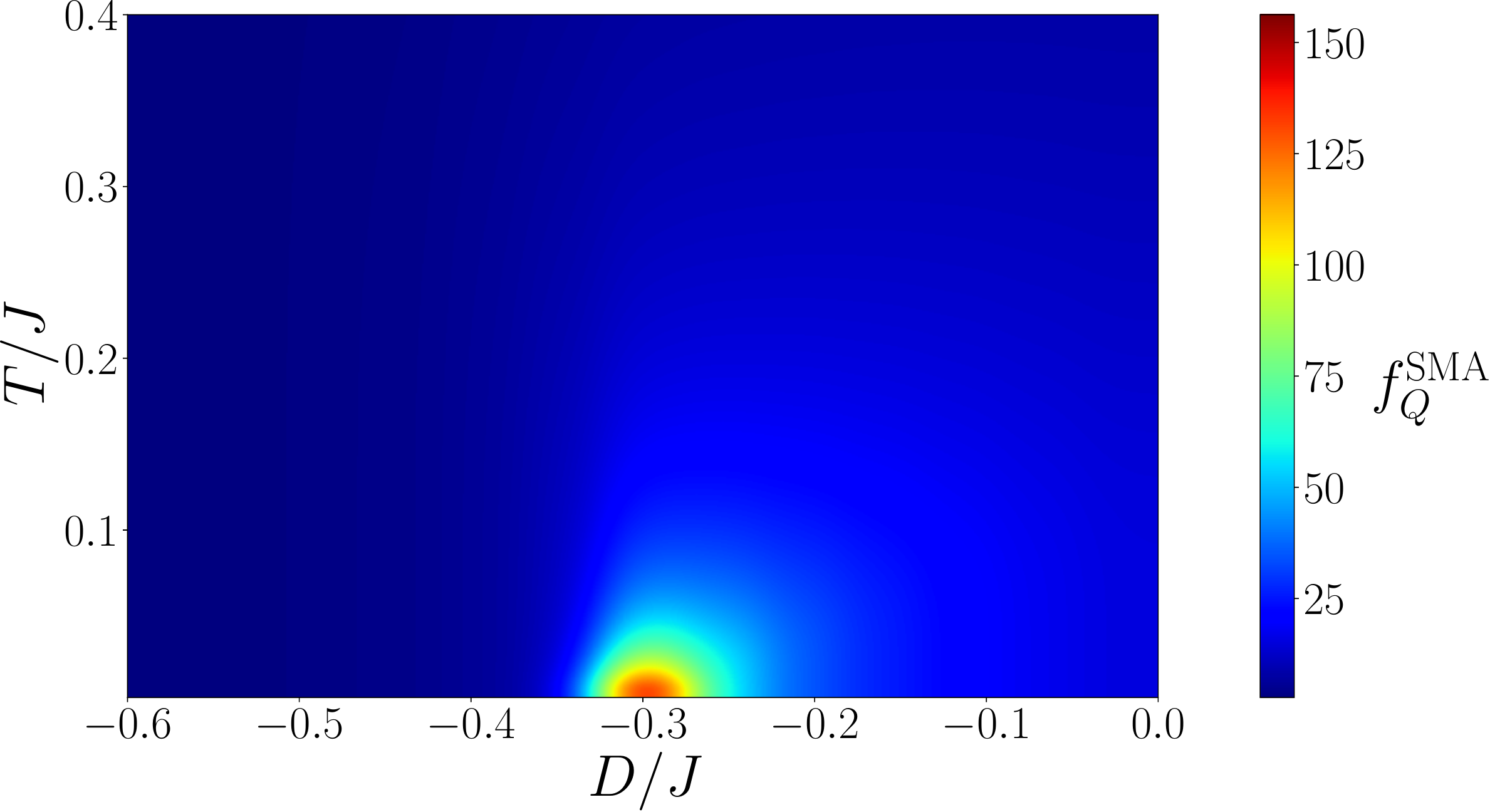}
	\caption{QFI density detected for $\hat{O}=\sum_r e^{ikr}S_r^z$ around the phase transition from the isotropic Haldane phase to the anti ferromagnetic phase for N=256. Obtained from SSE
	 results ($\beta=400$) and periodic boundary conditions} 
	\label{fig:qfipd}
\end{figure}
We use stochastic series expansion~\cite{sandvik1991quantum,sandvik1999stochastic,Syljuaasen2002} (SSE) techqniques to numerically study the QFI within the single mode approximation.
All of the SSE simulations used in this section use on the order of $10^6$ Monte Carlo sweeps. 
The data for each obersvable is binned into groups of $1000$ with the error bars estimated by taking the average variance over the bins. 

In order to examine the quantum Fisher information we consider the
operator, $\hat{O} = \sum_r e^{ikr}\hat{S}_r^z$. The equal time structure
factor for this operator corresponds to the spectrum of spontaneous
fluctuations in the longitudinal channel. Using Eq.~(\ref{disp}) we may compute
the bound on the dispersion for the heavy magnon to be \cite{so1994equal}, 
\begin{equation}
	\omega_{\text{SMA}} = \frac{J\left(C_{xx}^{r, r+1} + C_{yy}^{r, r+1}\right)\left(1-\cos(k)\right)}{S_0(k)},
	\label{wsma}
\end{equation}
where $C_{\alpha\beta}^{ij} \coloneqq \qexp{\hat{S}_i^\alpha \hat{S}_j^\beta}$. 
For the case of periodic boundary conditions the ground state is not
degenerate. We are here concerned with the singlet heavy magnon state. In this
case $\eta_0=\eta_1=1$ and thus, as per equation (\ref{eq:factor}), the leading thermal correction does not effect the QFI.
Since the single mode
approximation assumes these contributions to be small we ignore these thermal
corrections.  We may now apply the
single-mode approximation to compute the QFI density,
\begin{equation}
  f_Q(k) = 4\tanh^2\left(\frac{\omega_k}{2T}\right)S_0(k) + \int_{-\infty}^\infty\dd\omega\tanh^2\left(\frac{\omega}		{2T}\right)\tilde{S}(\omega, k),
\end{equation}
where we shall neglect the last term arising from the continuum contribution. 
Since this last term corresponds to a positive contribution we would expect to obtain a lower bound on the QFI. 
We argue, however, that the dominant effect, particularly near the isotropic point, will come from the inequality, $\omega\leq\omega_{SMA}$, Eq.~(\ref{disp}). Hence, we believe that an overestimation of the QFI density is the more
likely scenario. However, we expect this approximation to be rather good at low temperatures close to the isotropic point, $D=0$,
where we then obtain the estimate for $f_Q$,
\begin{equation}
  f^{\text{SMA}}_Q(k) \sim 4\tanh^2\left(\frac{\omega_{\text{SMA}}}{2T}\right)S_0(k).
	\label{eq:fqsma}
\end{equation}
We also note that the main
$T$-dependence of $f^{\text{SMA}}_Q(k)$ is now through the argument of the $\tanh$.
\begin{figure}[t!]
	\includegraphics[width=0.5\textwidth]{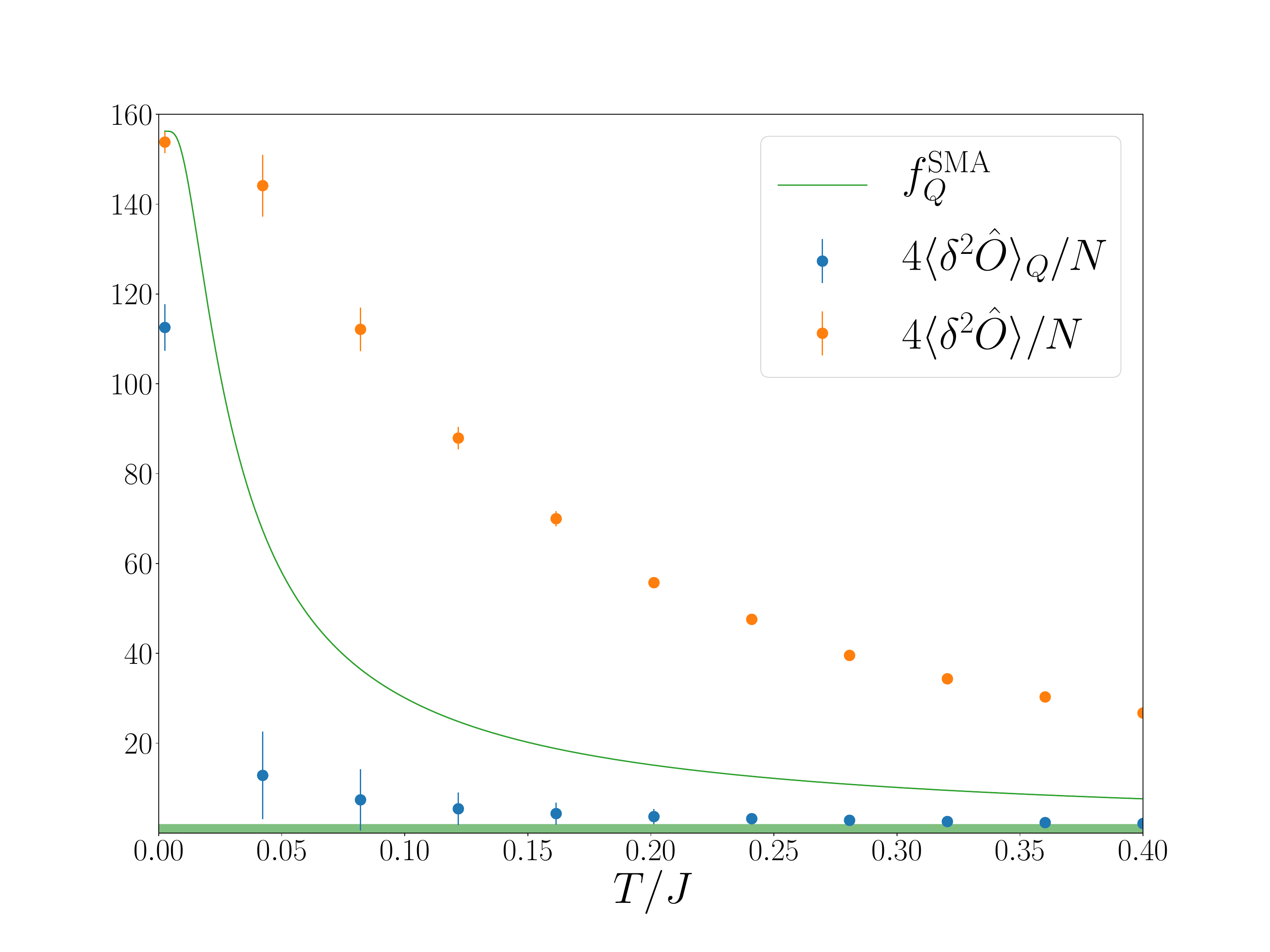}
	\caption{Finite temperature QFI density and QV above the critical point for $N=256$.
        $f_Q^{\text{SMA}}$ is obtained from simulations at $\beta=400$.
Upper and lower bounds for $f_Q$ given by $4\langle\delta^2\hat{O}\rangle/N$ and $4\langle\delta^2\hat{O}\rangle_Q/N$ are also shown for a range of temperatures.
        The green shaded region indicates the threshold $f_Q$ has to exceed for bi-partite entanglement to be present.}
	\label{fig:crittemp}
\end{figure}

Near the critical point it is expected that the continuum will contribute more
significantly to the behaviour of the system especially at non-zero
temperature, since the excitation gap closes. The finite temperature results at
the critical point are therefore less reliable than those at the isotropic
point, where the population is mainly concentrated in the ground state until
temperatures of the order of half the gap ($T\sim 0.2 J$) are reached.
With anti-ferromagnetic exchange, the equal time structure factor peaks at the
$k=\pi$ mode (Fig. \ref{fig:isochar})(a). Thus the quantum Fisher information is
maximal at the edge of the first Brillouin zone as shown in Fig.~\ref{fig:isochar}(b)
where $f^{\text{SMA}}_Q(k)$ is shown throughout Brillouin zone.
This corresponds to parameterizing the path through the space of density matrices using the
staggered magnetization. At $k=0$, $\hat{O}$ becomes the total magnetization
which commutes with the Hamiltonian and thus cannot detect entanglement. As we
approach $k=0$ the single mode approximation also becomes invalid since it is known
that the well-defined single mode present around $k=\pi$ merges into the continuum.
Fortunately the behaviour of the single mode approximation remains well
controlled at the edge of the Brillouin zone where the QFI density detected by
momentum space magnetization is maximal. In the following we therefore exclusively
focus on $k=\pi$.

Let us first consider the finite temperature behaviour of the entanglement at
the isotropic point, $D=0$. Using the QV we have established a lower bound on the
QFI density. On the other hand, there is a well-known upper bound~\cite{Toth2014} on $f_Q$ as well: $f_Q\leq 4\langle \delta^2\hat{O}\rangle/N$, 
where $\langle \delta^2\hat{O}\rangle$ refers to the total variance, Eq.~(\ref{eq:Vartot}).
Combining this with Eq.~(\ref{eq:fqsma}) we then obtain: 
\begin{equation}
	4\langle\delta^2\hat{O}\rangle_Q \leq F_Q \sim F_Q^{\text{SMA}}\leq 4\langle\delta^2\hat{O}\rangle.
\end{equation}
In Fig.~\ref{fig:isotemp} are shown results for
$f^{\text{SMA}}_Q(k=\pi)$ for a range of temperatures. 
In this regime the single-mode approximation should work quite well up until
approximately half the gap. 
Indeed, $f_Q^\text{SMA}$ is clearly within the hard upper and lower bounds on $f_Q$ given
by $4\langle\delta^2\hat{O}\rangle/N$ and $4\langle\delta^2\hat{O}\rangle_Q/N$ for all temperatures shown.
We see that up until this point the approximated
QFI density predicts the presence of multipartite entanglement well into this
regime. If we use the quantum variance as a lower bound on $f_Q(k)$ then it
predicts multipartite entanglement to temperatures
approaching the gap. In Fig.~\ref{fig:isotemp} the shaded green region indicates
the threshold to be exceeded for bi-partite to be present and we note that both
estimates of $f_Q(k=\pi)$ indicate the presence of bi-partite entanglement up to
temperatures close to the gap.

We may now ask how the ground state QFI density will behave as we
approach the quantum critical point. Fig.~(\ref{fig:qfipd}) show the QFI density
indicated by the color intensity for a range of temperatures and $D$ values 
for a system size of 256 and $\beta=400$ with periodic boundary conditions. We see that the QFI density is divergent at the
quantum critical point, as expected from the behaviour of $\hat{O}$.
Fig.~(\ref{fig:crittemp}) clearly shows the
The QFI density predicted by the single mode
approximation decays rapidly above the critical point, as the gap has now
effectively closed. Perhaps surprisingly,
 $f_Q^\text{SMA}$ is also here clearly withing the hard upper and lower bounds on $f_Q$ given
by $4\langle\delta^2\hat{O}\rangle/N$ and $4\langle\delta^2\hat{O}\rangle_Q/N$ for all temperatures shown.
We see in this case that
there is still persistent multipartite entanglement at finite temperatures
above the quantum critical point. 
\begin{figure}[t!]
	\includegraphics[width=0.48\textwidth]{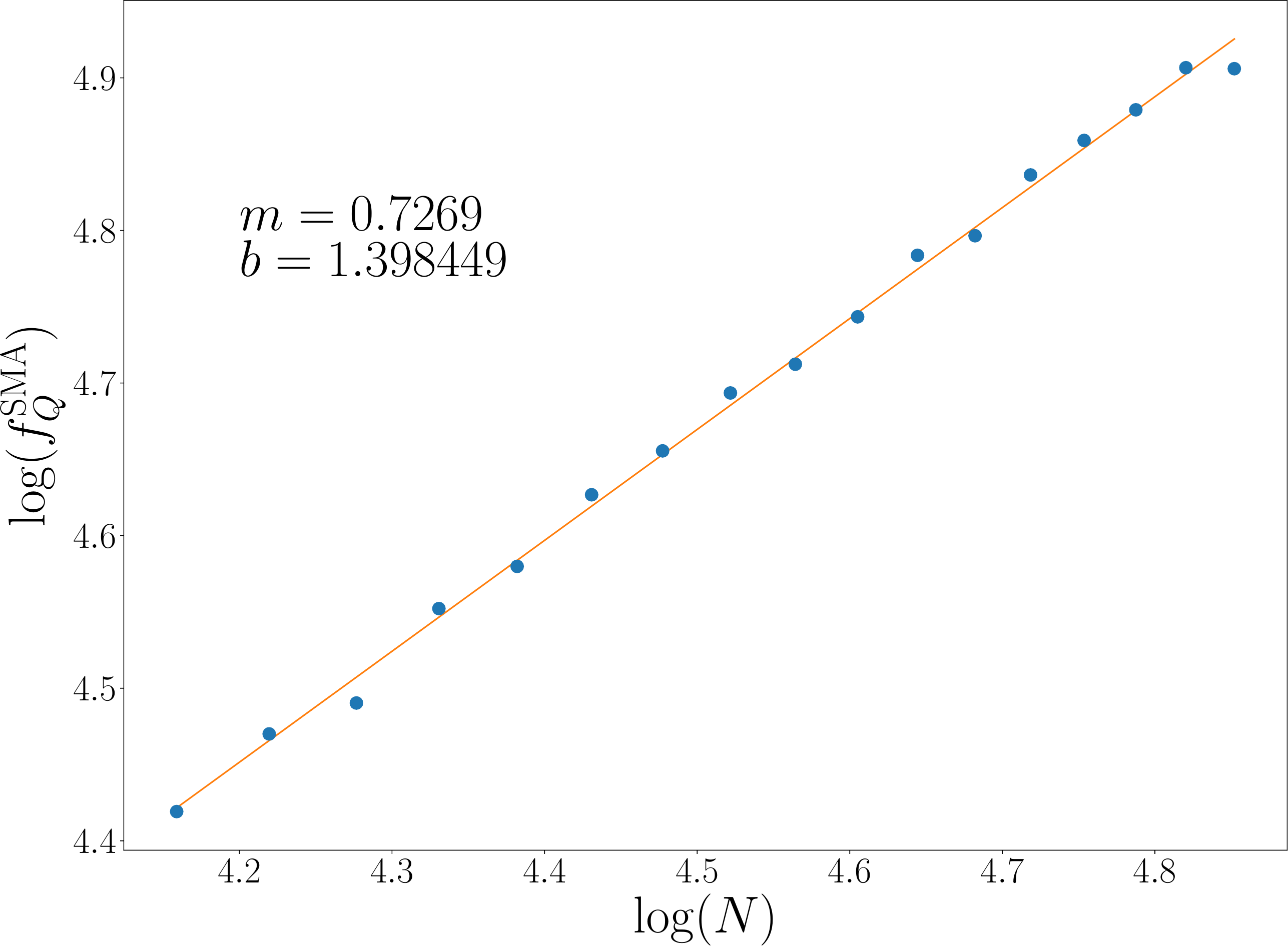}
        \caption{Finite size scaling of the QFI density with system size for
        even system sizes between $N=64$ and $N=128$. Scaling was performed at
        $D=D_C=-0.31$ at $\beta=400$ small enough that the system size would be the
        relevant perturbation to the scaling. The critical exponent with error
        due to the fit is found to be $\Delta_Q = 0.7269(1)$ with purely statistical error estimate.}\label{fig:QFIscaling}
\end{figure}

The divergence of the entanglement at the critical point is seen by examining
the QFI density for various systems sizes. Fig. \ref{fig:QFIscaling} demonstrate
the divergent scaling of both the QFI and the QV. Due to the fact that the
Haldane-N\'{e}el transition is in the Ising universality class we can compute
theoretically what the finite size scaling of the QFI density at the critical
point must be. At low enough temperatures this is for finite systems given by Eq.~(\ref{eq:fscaling}).
For the Ising universality class the critical exponent for the
staggered magnetization is given by $\Delta_\alpha = 1/8$. This is confirmed in
\cite{albuquerque2009quantum} using cluster expansion methods. This should give
a QFI density scaling of $\Delta_Q=3/4$. By examining even system sizes between
$N=64$ and $N=128$ at a $\beta=400$ at a value of $D_C^{HN}=-0.31$ we estimate a QFI
density scaling of $\Delta_Q = 0.7269(1)$. The error quoted here is associated
with the quality of the linear regression. It does not account for systematic
errors in the measurement of the QFI density. In order to estimate these
systematic errors we examine subsets of four points and determine the
maximum and minimum slopes that could be inferred from such four point subset of
the data. Using this we estimate a deviation of at least $\pm 0.06$. Thus
the estimated scaling is $\Delta_Q = 0.73 \pm 0.06$. This estimate is consistent
with the Ising universality class predicted for the Haldane-N\'{e}el transition which
is surprising since we would not expect the single-mode approximation to give
reliable results at the quantum critical point.

\section{Conclusions}~\label{sec:conclusions}
Using the single mode approximation we have shown that it is possible to obtain a quite
simple estimate of the QFI density that should yield reliable results at temperatures
well below the gap. We studied the $S=1$ anti-ferromagnetic spin chain with uniaxial anisotropy
within this approximation.
The approximation yields results that are within rigoruous upper and lower bounds for all tempertures studied.
Clear signatures of multipartite entanglement were found at the isotropic point, $D=0$
with the QFI density diverging when approaching the quantum critical point.
When combined with the QV, the single-mode approximated QFI allows one to place both upper
and lower bounds on the finite temperature entanglement of gapped systems. More precise techniques
for calculating the QFI density at finite temperatures in strongly correlated systems would clearly
be very desirable. Alternatively, sharper lower or upper bounds on the QFI density than we have discussed
here would be very valuable.

We also note that the QFI has been linked to the canonical energy in gravitational physics~\cite{Lashkari2016} and
can be expressed in terms of the relative entropy~\cite{Raamsdonk2016}, developments which could potentially be exploited
for more efficient numerical calculations of the QFI.

\section{Acknowledgements}
This research was supported by NSERC and
enabled in part by support provided by (SHARCNET) (www.sharcnet.ca) and Compute/Calcul Canada (www.computecanada.ca).


\bibliography{bibtex}

\end{document}